\documentclass[twocolumn,secnumarabic,amssymb, nobibnotes, aps, prd]{revtex4-2}

\usepackage{dcolumn}
\usepackage{bm} 
\usepackage{graphicx} 
\usepackage{color}
\usepackage{subfigure}
\usepackage{hyperref}
\usepackage{latexsym}
\usepackage{amsthm}
\usepackage{amsmath}
\usepackage{amssymb}
\usepackage{verbatim}
\usepackage{wasysym}

\DeclareGraphicsExtensions{.jpg, .pdf, .mps, .png, .eps, .ps, .EPS, .gif}
\begin{document}
\def\be{\begin{equation}}
\def\ee{\end{equation}}

\def\bc{\begin{center}}
\def\ec{\end{center}}
\def\bea{\begin{eqnarray}}
\def\eea{\end{eqnarray}}

\newcommand{\avg}[1]{\langle{#1}\rangle}
\newcommand{\Avg}[1]{\left\langle{#1}\right\rangle}

\newcommand{\cch}[1]{\left[#1\right]}
\newcommand{\chv}[1]{\left \{ #1\right \} }
\newcommand{\prt}[1]{\left(#1\right)}
\newcommand{\aver}[1]{\left\langle #1 \right\rangle}
\newcommand{\abs}[1]{\left| #1 \right|}

\def\ie{\textit{i.e.}}
\def\etal{\textit{et al.}}
\def\m{\vec{m}}
\def\G{\mathcal{G}}
\def\fig{FIG.}
\def\tab{TABLE}

\newcommand{\gin}[1]{{\bf\color{magenta}#1}}
\newcommand{\bobred}[1]{{\bf\color{red}#1}}
\newcommand{\bobredx}[1]{#1}
\newcommand{\bobz}[1]{{\bf\color{magenta}#1}}

\title{Extended-range percolation in five dimensions}

\author{Zhipeng Xun$^a$}
\email{zpxun@cumt.edu.cn}
\author{Dapeng Hao$^{a,b}$}
\email{dphao@cumt.edu.cn}
\author{Robert M. Ziff$^b$}
\email{rziff@umich.edu}
\affiliation{$^a$ School of Material Sciences and Physics, China University of Mining and Technology, Xuzhou 221116, China}
\affiliation{$^b$ Center for the Study of Complex System and Department of Chemical Engineering, University of Michigan, Ann Arbor, Michigan 48109-2800, USA}

\date{\today}

\begin{abstract}
Percolation on a five-dimensional simple hypercubic (\textsc{sc(5)}) lattice with extended neighborhoods is investigated by means of extensive Monte Carlo simulations, using an effective single-cluster growth algorithm. The critical exponents, including $\tau$ and $\Omega$, the asymptotic behavior of the threshold $p_c$ and its dependence on coordination number $z$ are investigated. Using the bond and site percolation thresholds $p_c = 0.11817145(3)$ and $0.14079633(4)$ respectively given by Mertens and Moore [Phys.\ Rev.\ E \textbf{98}, 022120 (2018)], we find critical exponents of $\tau = 2.4177(3)$, $\Omega = 0.27(2)$ through a self-consistent process. The value of $\tau$ compares favorably with a recent five-loop renormalization predictions $2.4175(2)$ by Borinsky et al.\ [Phys.\ Rev.\ D \textbf{103}, 116024 (2021)], the value  2.4180(6) that follows from the work of Zhang et al.\ [Physica A \textbf{580}, 126124 (2021)], and the measurement of $2.419(1)$ by Mertens and Moore. We also confirmed the bond threshold, finding $p_c = 0.11817150(5)$.   \textsc{sc(5)} lattices with extended neighborhoods up to 7th nearest neighbors are studied for both bond and site percolation. Employing the values of $\tau$ and $\Omega$ mentioned above, thresholds are found to high precision. For bond percolation, the asymptotic value of $zp_c$ tends to Bethe-lattice behavior ($z p_c \sim 1$), and the finite-$z$ correction is found to be consistent with both  and $zp_{c} - 1 \sim a_1 z^{-0.88}$ and $zp_{c} - 1 \sim a_0(3 + \ln z)/z$.  For site percolation, the asymptotic analysis is close to the predicted behavior $zp_c \sim 32\eta_c = 1.742(2)$ for large $z$, where $\eta_c = 0.05443(7)$ is the continuum percolation threshold of five-dimensional hyperspheres given by Torquato and Jiao [J. Chem.\ Phys.\ \textbf{137}, 074106 (2015)]; finite-$z$ corrections are accounted for by taking $p_c \approx c/(z + b)$ with $c=1.722(7)$ and $b=1$.  
\end{abstract}


\maketitle
\section{Introduction}
Percolation \cite{BroadbentHammersley1957,StaufferAharony1994}, which has a history of at least 70 years, has numerous applications in various branches of science, including statistical physics where it plays a central paradigmatic role \cite{StaufferAharony1994,BolandtabaSkauge2011,MourzenkoThovertAdler2011,Henley1993,GuisoniLoscarAlbano2011,RahmanHassan19,MooreNewman2000,ScullardJacobsen2020,Ziff2021}. The percolation transition, one of its central concepts, occurs when an infinite cluster spanning over the entire system appears, with a statistically independent critical site or bond probability $p_c$, which is called the percolation threshold. 

In order to investigate percolation systematically, researchers have established many kinds of percolation models, including site and bond percolation on regular and random lattices, continuum percolation \cite{XuWangHuDeng2021,MertensMoore2012,QuintanillaZiff2007,TarasevichEserkepov20}, correlated percolation \cite{Kantor1986,ZierenbergFrickeMarenzSpitznerBlavatskaJanke2017}, bootstrap percolation \cite{ChoiYu2020,ChoiYu2019,MuroBuldyrevBraunstein2020,MuroValdezStanleyBuldyrevBraunstein2019}, and directed percolation \cite{WangZhouLiuGaroniDeng13,Grassberger2009,Grassberger2009b,Jensen2004}. Among these models, percolation on lattices with extended neighborhoods has been of longstanding interest due to its close links to several important topics, including problems of adsorption of extended shapes on a lattice \cite{KozaKondratSuszcaynski2014,KozaPola2016}, small-world networks \cite{Kleinberg2000}, and the spread of epidemics \cite{SanderWarrenSokolov2003,Ziff2021}. 

Since the  concept of the ``equivalent neighbor model" was introduced by Dalton, Domb and Sykes in 1964 \cite{DaltonDombSykes64,DombDalton1966,Domb72}, numerous investigations of percolation on lattices with extended neighborhoods, including compact regions in a diamond shape on a square lattice \cite{GoukerFamily83}, \textsc{bcc} and \textsc{fcc} lattices \cite{JerauldScrivenDavis1984,GawronCieplak91}, the eleven Archimedean lattices (``mosaics") \cite{dIribarneRasigniRasigni95,dIribarneRasigniRasigni99,dIribarneRasigniRasigni99b}, lattices with combinations of ``complex neighborhoods" in various dimensions \cite{MalarzGalam05,GalamMalarz05,MajewskiMalarz2007,KurzawskiMalarz2012,Malarz2015,KotwicaGronekMalarz19,Malarz2020,Malarz21,Malarz23},
overlapping shapes on a lattice \cite{KozaKondratSuszcaynski2014,KozaPola2016}, and distorted lattices \cite{MitraSahaSensharma22,MitraSensharma23} have been carried out. While much of the earlier work concerned site percolation, bond percolation on extended lattices has also been studied more recently \cite{OuyangDengBlote2018,DengOuyangBlote2019,XunZiff2020,XunZiff2020b,XuWangHuDeng2021,XunHaoZiff2022,ZhaoYanXunHaoZiff2022}. Extended-range percolation is also referred to as medium-range percolation \cite{DengOuyangBlote2019}, range-$R$ percolation \cite{FreiPerkins2016,Hong21}, complex-neighborhood percolation \cite{MajewskiMalarz2007}, and long-range percolation \cite{GoukerFamily83,dIribarneRasigniRasigni99b}.


Many studies focus on exploring the correlations between $p_{c}$ and properties of lattices, especially the coordination number $z$. For bond percolation, one expects that for large $z$,  Bethe-lattice behavior
\begin{equation}
    p_c = \frac{1}{z-1}
\label{eq:zpcbond}
\end{equation}
will hold, because for large $z$,  $p$ will be small, and the chance of a link to  hit the same site twice is low.  Consequently, the system behaves basically like a tree \cite{Penrose93}.  With regard to finite-$z$ corrections, theoretical analysis for bond thresholds has recently been given by Frei and Perkins \cite{FreiPerkins2016}, Hong \cite{Hong21}, and  Xu et al.\ \cite{XuWangHuDeng2021} as
\begin{equation}
    zp_{c} - 1 \sim a_{0} \frac{\ln z}{z}
\label{eq:lnz}
\end{equation}
for $d = 4$, and 
\begin{equation}
    zp_{c} - 1 \sim a_{1}z^{-x}
\label{eq:zpcbond2}
\end{equation}
for $d=2$ and 3, where $x = (d-1)/d$. 

For site percolation, it has been argued \cite{Domb72,dIribarneRasigniRasigni99b,KozaKondratSuszcaynski2014,KozaPola2016} that the threshold $p_c$ for large $z$ can be related to the continuum percolation threshold $\eta_c$ for objects of the same shape as the neighborhood, and this relationship is further clarified \cite{Penrose93,XunHaoZiff2021} by
\begin{equation}
    zp_c = 2^{d} \eta_{c}
\label{eq:zpcsite}
\end{equation}
for large $z$.
In Ref.\ \cite{XunHaoZiff2021}, we found that finite-$z$ corrections can be well taken into account by assuming 
\begin{equation}
p_c = \frac{c}{z+b}
\label{eq:zpcsite2}
\end{equation}
where $b$ and $c$ are constants. This relation can also be written as  
$z = c/p_c - b$.

Many important results have been achieved in two, three and four dimensions. However, there are still many gaps in five dimensions. In fact, we cannot neglect the investigation of five dimensions. On one hand, five dimension is close enough to the upper critical dimension of six for $\epsilon = 6 - d$ series analysis to have the possibility of yielding precise results. On the other hand, it is also interesting to know the dependence of the thresholds and critical exponents upon the dimensionality. These motivate the research of this work.

In this paper, by employing an effective single-cluster growth algorithm, bond and site percolation on five-dimensional simple hypercubic (\textsc{sc(5)}) lattices are simulated. Using the existing bond and site percolation thresholds given by Mertens and Moore \cite{MertensMoore2018}, the critical exponents  
$\tau$ and $\Omega$ are estimated and are found to be consistent with previous work.
Then, using these $\tau$ and $\Omega$, numerical simulations of the \textsc{sc(5)} lattice with extended neighborhoods are carried out. We find precise bond and site percolation thresholds for \textsc{sc(5)} lattices with up to 7th nearest neighbors, and investigate the asymptotic behavior of the threshold $p_c$ and its dependence on coordination number $z$. We note that studies of percolation in five dimensions go back at least to the 1980's \cite{JanHongStanley85,KnackstedtMcCraryPayandehRoberts88}.


The remainder of the paper is organized as follows. Section \ref{sec:single} describes the single-cluster growth algorithm and contains a brief theoretical background. The results and analysis for critical exponents, bond percolation thresholds and site percolation thresholds are given in Secs.\ \ref{sec:critical}, \ref{sec:bond} and \ref{sec:site}, respectively, and in Sec.\ \ref{sec:conclusions} we present our conclusions. 

\section{Single-cluster growth algorithm and theoretical background}
\label{sec:single}
In our Monte Carlo simulations, a single-cluster growth algorithm is employed. From a seeded site on the lattice, a large number of individual clusters are generated. Defining $s$ as the size of the clusters in terms of the number of connected sites, we put clusters with different sizes in bins with the intervals $[2^n, 2^{n+1})$ for $n=0,1,2,\ldots$. Clusters still growing when they reach an upper size cutoff $s_\mathrm{max}=2^{n_\mathrm{max}}$, which is set in advance to avoid wrapping around the boundaries and also to limit the run time, are counted in the last bin $n=n_\mathrm{max}$. Numerous simulation runs are carried out for each lattice.  From the data in the bins we can extract the distribution of the clusters grown, characterized by $P_{\ge s}$, the probability that a vertex is connected to a cluster of size greater than or equal to $s$.  This quantity is related to $n_s(p)$, the number of clusters (per site) containing $s$ occupied sites as a function of the site or bond occupation probability $p$ on a fully populated lattice as discussed below.

Near the percolation threshold $p_c$, $n_s$ is expected to behave as \cite{LorenzZiff1998}
\begin{equation}
    n_{s} \sim A_0 s^{-\tau} (1 + B_0 s^{-\Omega} + C_0 (p-p_c)s^\sigma\dots),
    \label{eq:ns}
\end{equation}
where the critical exponent $\sigma$, the Fisher exponent $\tau$, and leading correction-to-scaling exponent $\Omega$ are expected to be universal (having same value for all lattices and percolation types in a given dimension), while $A_0$ and $B_0$ are constants that depend upon the system and are thus non-universal. Then the probability a vertex belongs to a cluster with size greater than or equal to $s$ will be given by
\begin{equation}
P_{\geq s} = \sum_{s'=s}^\infty s' n_{s'} \sim A_1s^{2-\tau} (1+B_1 s^{-\Omega} + C_1(p-p_c) s^\sigma + \ldots),
\label{ps}
\end{equation}
where $A_1 = A_0/(\tau-2)$, $B_1 = (\tau-2)B_0/(\tau+\Omega-2)$, and $C_1 = (\tau-2)C_0/(\tau-\sigma-2)$. Multiplying both sides of Eq.\ (\ref{ps}) by $s^{\tau-2}$ yields
\cite{LorenzZiff1998}
\begin{equation}
s^{\tau-2}P_{\geq s} \sim A_1 (1+B_1 s^{-\Omega}+C_1 (p-p_c)s^\sigma+\dots),
\label{staup}
\end{equation}
At $p = p_c$, Eq.\ (\ref{ps}) implies that 
\cite{ZiffBabalievski99,XunZiff2020}
\begin{equation}
Q(s) \equiv \frac{\ln P_{\geq 2s} - \ln P_{\geq s}}{\ln 2} \sim (2 - \tau) + B_2 s^{-\Omega}
\label{localslope}
\end{equation}
where $B_2 = B_1(2^{-\Omega}-1)/\ln 2$.  The quantity $Q(s)$ represents the local slope between pairs of points with $s$ and $2s$ on a plot of $\ln P_{\geq s}$ versus $\ln s$.

Eqs.\ (\ref{staup}) and (\ref{localslope}) imply that, at $p = p_c$, there should be linear relationships if we make plots of $s^{\tau-2}P_{\geq s}$ versus $s^{-\Omega}$, and $Q(s)$ versus $s^{-\Omega}$, for large $s$, where Eq.\ (\ref{localslope}) also shows that the $y$-intercept of the straight line will give the value of ($2 - \tau$). These relations provide a self-consistent method to estimate the values of $p_c$, $\tau$, and $\Omega$ at the same time. 
One can see Refs. \cite{LorenzZiff1998,XunZiff2020,XunZiff2020b} for more details about the single-cluster growth method and its theory.

We use the notation $\textsc{sc(5)-}a,b,...$ to indicate a five-dimensional simple hypercubic lattice with the $a$-th nearest neighbors, $b$-th nearest neighbors, etc. Monte Carlo simulations are carried out on systems of size $L\times L \times L \times L \times L$ with $L = 64$ under periodic boundary conditions. Many independent samples are produced; from these simulations $P_{\ge s}$ are found, and the quantities $s^{\tau-2}P_{\geq s}$ and $Q(s)$ could easily be calculated.   The four-tap R(471,1586,6988,9689) pseudo-random number generator  \cite{Ziff98} (GNU gsl\_rng\_gfsr4) was used.

\section{Critical exponents}
\label{sec:critical}
For bond and site percolation on the \textsc{sc(5)} lattice, there exist  several works focused on finding the thresholds \cite{Kirkpatrick1976, GauntSykesRuskin1976, Grassberger2003, DammerHinrichsen2004, KozaPola2016}. At present, the most precise values are $p_c = 0.11817145(3)$ for bond percolation and $p_c = 0.14079633(4)$ for site percolation, given by Mertens and Moore \cite{MertensMoore2018}. In order to independently determine the values of $\tau$ and $\Omega$, we use these thresholds, which are sufficiently precise for our study. The upper size cutoffs are set to be $s_\mathrm{max} = 2^{16}$ occupied sites for both bond and site percolation, and the number of runs are $2 \times 10^{10}$ for bond percolation and $10^{10}$ for site percolation.  We checked that with this cutoff, clusters did not see themselves after wrapping around the system, which was found to happen only when the cutoff was increased to $2^{18}$.

\begin{figure}[htbp] 
\centering
\includegraphics[width=2.9in]{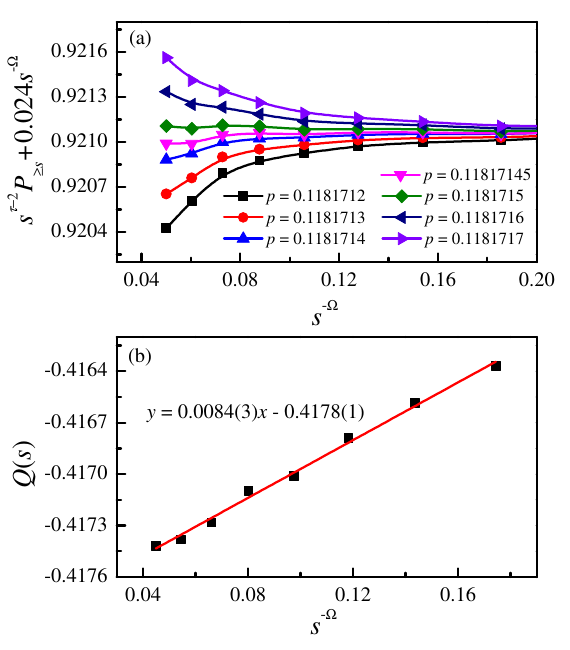}
\caption{Plots of $s^{\tau-2}P_{\geq s}$ with $\tau = 2.4178$ (a) and $Q(s)$ (b) versus $s^{-\Omega}$ with $\Omega = 0.28$ for bond percolation of the \textsc{sc(5)} lattice under different values of $p$ (a), and for $p = 0.11817145$ (b). In (a), we added the linear function $0.024 s^{-\Omega}$ in order to make the slope on the right nearly horizontal and allow us to expand the vertical scale and separate the curves.  For $p \ne p_c$, the data in (a) behaves as $0.921+C_1(p-p_c)(s^{-\Omega})^{-\sigma/\Omega}$ with $\sigma/\Omega\approx -1.83$. Data fitting shows that these values of $p_c$, $\tau$ and $\Omega$ are self-consistent.}
\label{fig:bond}
\end{figure}

Figure \ref{fig:bond} shows the relations of $s^{\tau-2}P_{\geq s}$ versus $s^{-\Omega}$ and $Q(s)$ versus $s^{-\Omega}$ under $p = 0.11817145$. To check the value of $p_c$ for bond percolation, we also simulated $s^{\tau-2}P_{\geq s}$ versus $s^{-\Omega}$ under occupying probabilities $p = 0.1181712, \ldots, 0.1181717$, as shown in Fig.\ \ref{fig:bond}(a). A large number of runs, $2\times 10^{9}$, were required to distinguish the behavior for these very close values of $p$.  The threshold is determined by the value of $p$ where the plot is linear for larger $s$.  The resulting estimate of the threshold $p_c = 0.11817150(5)$  is consistent with previous results as shown in table \ref{tab:pcbondSC5}. 
Overall, we found that self-consistent results can be found if we choose the values of $p = 0.11817145$, $\tau = 2.4178$ and $\Omega = 0.27$, 
that is, both $s^{\tau-2}P_{\geq s}$ versus $s^{-\Omega}$ and $Q(s)$ versus $s^{-\Omega}$ show linear behavior, meanwhile the intercept tends to $-0.4178(1)$ for ($2 - \tau$), as shown in Fig.\ \ref{fig:bond}(b). In addition, the simulation results in Fig.\ \ref{fig:bond}(a) also indicate that if $p$ is away from $p_c$, the behavior will be non-linear.  

We note that the high-$d$ asymptotic series \cite{MertensMoore2018b} for bond thresholds predicts $p_c = 0.117410$; this series becomes more accurate as $d$ is increased.

\begin{table}[htb]
\caption{Values of the bond threshold for SC(5). MC = Monte Carlo.}
\begin{tabular}{lllll}
\hline\hline
    value                     &  year  & method & authors & reference  \\
    \hline
    0.11819(4)                &  1990  & Series & Adler et al. &\cite{AdlerMeirAharonyHarris90}\\
    0.118172(1)               &  2003  & MC & Grassberger &\cite{Grassberger2003}\\
    0.118172(1)               &  2004  & MC &Dammer \& Hinrichsen & \cite{DammerHinrichsen2004}\\
    0.11817145(3)             &  2018  & MC &Mertens \& Moore &\cite{MertensMoore2018}\\
    0.11817150(5)             &  & MC & this work   &   \\
\hline\hline
\end{tabular}
\label{tab:pcbondSC5}
\end{table}

\begin{figure}[htbp] 
\centering
\includegraphics[width=2.9in]{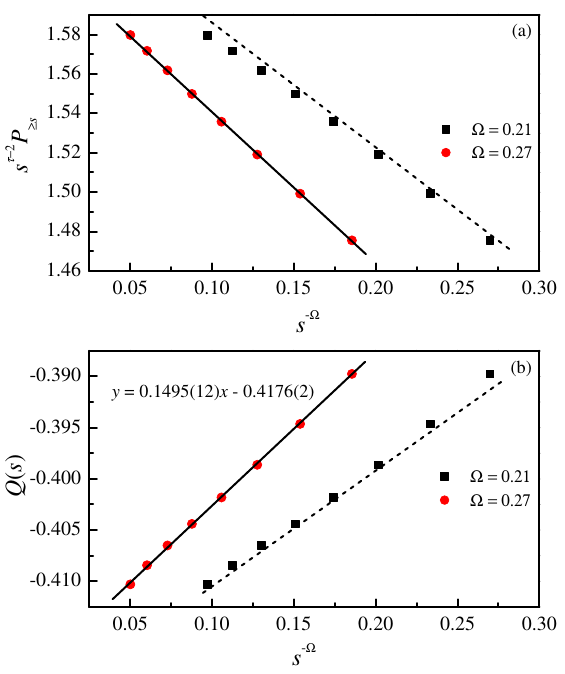}
\caption{Plots of $s^{\tau-2}P_{\geq s}$  with $\tau = 2.4176$ (a) and $Q(s)$ (b) versus $s^{-\Omega}$ with $\Omega = 0.21$ and $\Omega = 0.27$ for site percolation on the \textsc{sc(5)} lattice under $p = 0.14079633$. Clearly, better linear behavior obtains with $\Omega=0.27$ than $\Omega = 0.21$; dashed lines are included as guides to the eye.  Data fitting shows that these values of $p_c$, $\tau$ and $\Omega=0.27$ are self-consistent.}
\label{fig:site2}
\end{figure}

For site percolation, in Fig.\ \ref{fig:site2}, we exhibit the plots of $s^{\tau-2}P_{\geq s}$ versus $s^{-\Omega}$ and $Q(s)$ versus $s^{-\Omega}$ under the single value of $p = 0.14079633$. Self-consistent results can also be obtained when we choose the values of $\tau = 2.4176$ and $\Omega = 0.27$.  We also show the behavior assuming $\Omega = 0.21$ \cite{BorinskyGraceyKompanietsSchnetz2021} and find inconsistent results for $\tau$ from the intercept of $Q(s)$ and non-linear behavior in both plots.

Comprehensively considering both bond and site percolation, we can estimate that the values of $\tau$ and $\Omega$ are
\begin{eqnarray}
\tau &=& 2.4177(3) \cr \cr 
\Omega &=& 0.27(2)
\label{eq:tauOmega}
\end{eqnarray}
where the numbers in parentheses represent errors in the last digit(s). Our value of $\tau$ is consistent with the previous  measurements, including the recent five-loop renormalization prediction by Borinsky et al., as shown in table \ref{tab:tauvalues}.
The value of $\Omega$ is somewhat higher than the value 0.210(2) of Borinsky et al., but consistent with an older and less precise series-study result 0.27(7) of Adler et al.\ \cite{AdlerMeirAharonyHarris90}.  There have been studies the exponent $y_1 = -\Omega d_f$ in other contexts than the size distribution (such as moments ratios and covariances), and the results are not definitive \cite{ZhangHouFangHuDeng21,TanDengJacobsen20}. It may be that larger systems and cutoffs are needed to find the true corrections-to-scaling behavior.

\begin{table}[htb]
\caption{Values of the critical exponent $\tau$ ($^*$using other exponents and scaling relations).}
\begin{tabular}{lllll}
\hline\hline
    value                     &  year  & method & authors & reference  \\
    \hline
    2.413(12)$^*$         &  1990  & Series & Adler et al. &\cite{AdlerMeirAharonyHarris90}\\
    2.412(4)             &  2001  & MC & Paul et al. &\cite{PaulZiffStanley2001}\\
    2.4171               &  2015  & Field Th. & Gracey & \cite{Gracey2015}\\
    2.419(1)             &  2018  & MC & Mertens \& Moore &\cite{MertensMoore2018}\\
    2.4180(6)$^*$           &  2021  & MC & Zhang et al.& \cite{ZhangHouFangHuDeng21,TanDengJacobsen20} \\
    2.4175(2)            &  2021  & Field Th. & Borinsky et al. &\cite{BorinskyGraceyKompanietsSchnetz2021}\\
    2.4177(3)             & &  MC & this work     & \\
\hline\hline
\end{tabular}
\label{tab:tauvalues}
\end{table}

\section{Bond percolation with extended neighborhoods}
\label{sec:bond}
For bond percolation, the lattices studied and runs carried out are listed in table \ref{tab:bondperholds}.  For each, the cutoff used was $s_\mathrm{max}=2^{16}$.

\begin{figure*}[htbp] 
\centering
\includegraphics[width=7in]{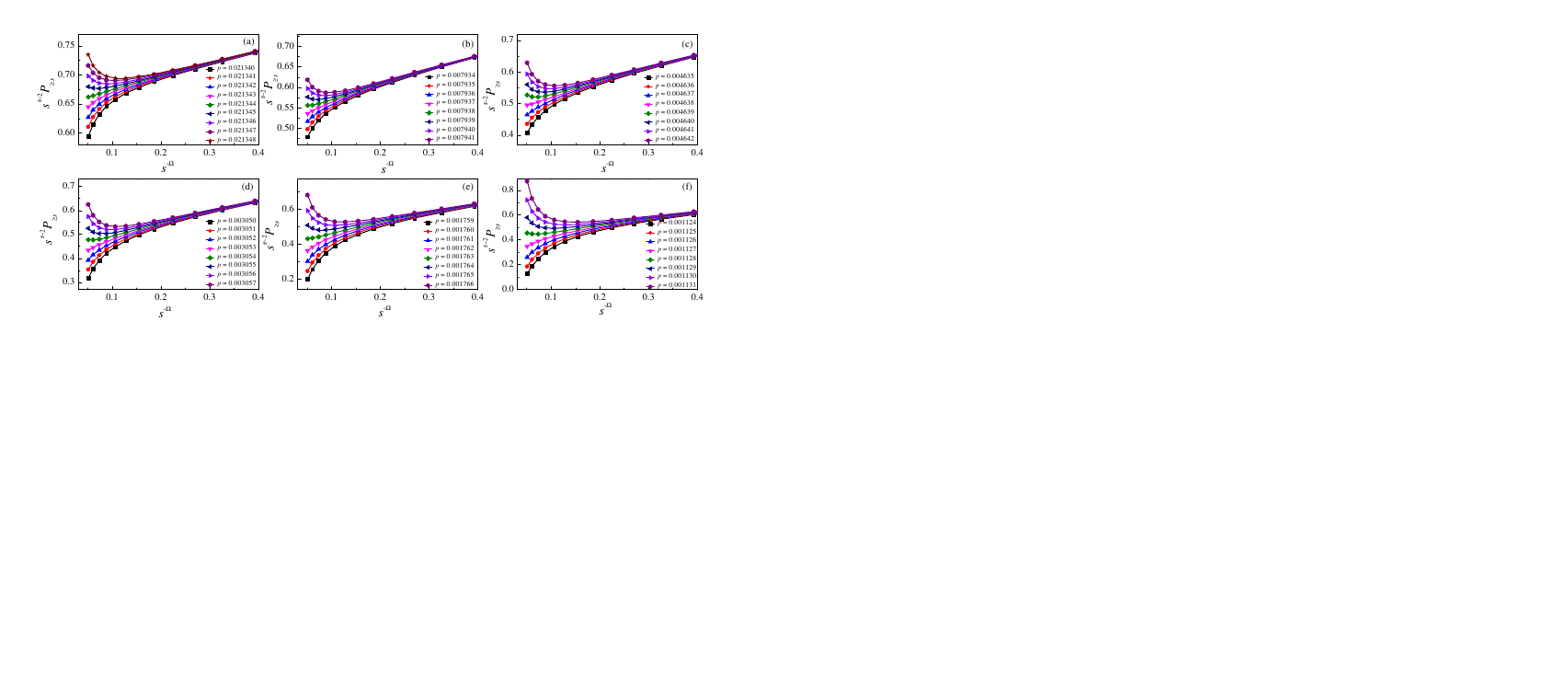}
\caption{Plot of $s^{\tau-2}P_{\geq s}$ versus $s^{-\Omega}$ with $\tau = 2.4177$ and $\Omega = 0.27$ for bond percolation of the \textsc{sc(5)}-1,2 (a), \textsc{sc(5)}-1,2,3 (b), \textsc{sc(5)}-1,2,3,4 (c), \textsc{sc(5)}-1,...,5  (d), \textsc{sc(5)}-1,...,6 (e), and \textsc{sc(5)}-1,...,7 (f)  lattices under different values of $p$.}
\label{fig:bondomega}
\end{figure*}

Using the values of the $\tau$ and $\Omega$ from Eq.\ (\ref{eq:tauOmega}), in Fig.\ \ref{fig:bondomega}, we plot $s^{\tau-2}P_{\geq s}$ versus $s^{-\Omega}$ for the
systems under different values of occupation probability $p$. Utilizing the linear property relation between $s^{\tau-2}P_{\geq s}$ and $s^{-\Omega}$ of Eq. (\ref{staup}), we identify the thresholds $p_c$ as summarized in table \ref{tab:bondperholds}. 

\begin{table}[htb]
\caption{Bond percolation thresholds of the \textsc{sc(5)}-1,...,$n$ lattices.  The coordination number $z$ is equal to the number of neighbors within maximum Euclidean distance $\sqrt{n}$.}
\begin{tabular}{ccrll}
\hline\hline
    lattice                     & runs   & $z$    & $p_{c}$         & $zp_{c}$  \\
    \hline
    \textsc{sc(5)-1=sc(5)}      & $2 \times 10^{9}$  & 10     & 0.11817150(5)   & 1.1817   \\    
    \textsc{sc(5)}-1,2          & $3 \times 10^{8}$  & 50     & 0.021348(2)     & 1.0674    \\
    \textsc{sc(5)}-1,2,3        & $3 \times 10^{8}$  & 130    & 0.0079375(5)    & 1.0319    \\
    \textsc{sc(5)}-1,2,3,4      & $3 \times 10^{8}$  & 220    & 0.004638(1)     & 1.0204    \\ 
    \textsc{sc(5)}-1,...,5      & $2 \times 10^{8}$  & 332    & 0.0030535(5)    & 1.0138    \\
    \textsc{sc(5)}-1,...,6      & $10^{8}$           & 572    & 0.001763(1)     & 1.0084    \\ 
    \textsc{sc(5)}-1,...,7      & $10^{8}$           & 892    & 0.0011275(5)    & 1.0057    \\
\hline\hline
\end{tabular}
\label{tab:bondperholds}
\end{table}

\begin{figure}[htbp] 
\centering
\includegraphics[width=2.9in]{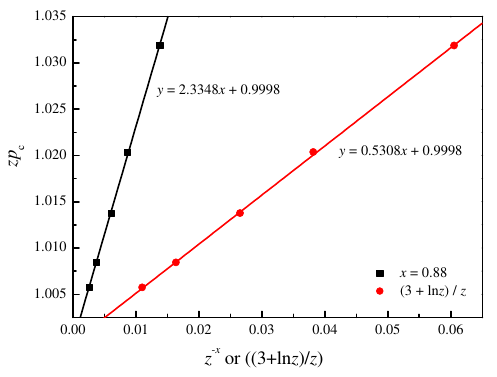}
\caption{Plot of $zp_c$ versus $z^{-x}$ and versus $(3 + \ln z)/z$ for bond percolation on the \textsc{sc}(5)-$1,...,n$ lattices with compact nearest neighborhoods for $n = 3,4,5,6,7$. If we pick $x=0.88$, we get an intercept close to 1, as well as a good linear fit. This data
can also evidently be fitted with $zp_c -1 \sim a_1(C + \ln z)/z$ with $C=3$.}
\label{fig:bondzpc2}
\end{figure}

For each lattice, the values of $zp_c$ are also shown in table \ref{tab:bondperholds}. It can be seen that $zp_c$ decreases and tends to the asymptotic value of $1$ with the increase of coordination number $z$. The relation of $zp_c$ versus $z^{-x}$, and also versus $(C + \ln z) / z$ is shown in Fig.\ \ref{fig:bondzpc2}. Data fitting indicates that when choosing $x = 0.88$, we achieve good linear behavior, meanwhile we get an intercept close to the asymptotic value of 1. There is a discrepancy between the result here and the situation of two and three dimensions where $x = (d-1)/d$, though the latter formula is only intended to apply for $d = 2$ and 3.   The data could also be fitted by $zp_c -1 \sim (C + \ln z)/z$. Compared to four dimensions of $zp_c -1 \sim (\ln z)/z$, here we need to add a higher-order correction term $C/z$ with $C = 3$.

\section{Site percolation with extended neighborhoods}
\label{sec:site}
For site percolation of \textsc{sc(5)} lattice with extended neighborhoods, the lattices considered and runs carried out are listed in table \ref{tab:siteperholds}. The cutoff for all was $s_\mathrm{max} = 2^{14}$. 

\begin{figure*}[htbp] 
\centering
\includegraphics[width=7in]{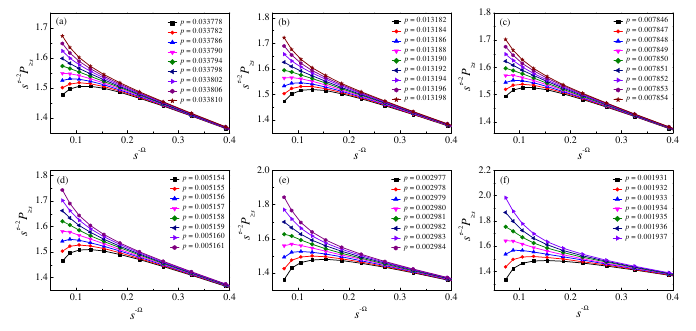}
\caption{Plots of $s^{\tau-2}P_{\geq s}$ versus $s^{-\Omega}$ with $\tau = 2.4177$ and $\Omega = 0.27$ for site percolation on the lattices \textsc{sc(5)}-1,2  (a), \textsc{sc(5)}-1,2,3 (b), \textsc{sc(5)}-1,2,3,4 (c), \textsc{sc(5)}-1,...,5 (d), \textsc{sc(5)}-1,...,6 (e), and \textsc{sc(5)}-1,...,7 (f) under different values of $p$.}
\label{fig:siteomega}
\end{figure*}

Similar to the bond percolation, the plots of $s^{\tau-2}P_{\geq s}$ versus $s^{-\Omega}$ for site percolation of the \textsc{sc(5)}-$1,...n$ lattices for $n = 2$ to 7,  under different values of occupation probability $p$, are shown in Fig.\ \ref{fig:siteomega}, and estimated thresholds for each lattice are summarized in table \ref{tab:siteperholds}. 

\begin{table}[htb]
\caption{Site percolation thresholds of the \textsc{sc(5)}-$1,...,n$ lattices.}
\begin{tabular}{ccrll}
\hline\hline
    lattice                     &   runs & $z$  & $p_{c}$          & $zp_{c}$  \\ \hline
    \textsc{sc(5)-1=sc(5)}    &    \cite{MertensMoore2018} & 10     & 0.14079633(4)      &    1.4080    \\
   \textsc{sc(5)}-1,2          &$3 \times 10^{8}$ & 50     & 0.033794(4)      & 1.6897    \\
    \textsc{sc(5)}-1,2,3        &$2 \times 10^{8}$& 130    & 0.013190(2)      & 1.7147    \\
    \textsc{sc(5)}-1,2,3,4      &$2 \times 10^{8}$& 220    & 0.007850(1)      & 1.7270    \\ 
    \textsc{sc(5)}-1,...,5      &$2 \times 10^{8}$& 332    & 0.0051575(5)     & 1.7123    \\
    \textsc{sc(5)}-1,...,6      &$10^{8}$& 572    & 0.0029805(5)     & 1.7048    \\ 
    \textsc{sc(5)}-1,...,7      &$10^{8}$& 892    & 0.001934(1)      & 1.7251    \\
\hline\hline
\end{tabular}
\label{tab:siteperholds}
\end{table}

\begin{figure}[htbp] 
\centering
\includegraphics[width=2.9in]{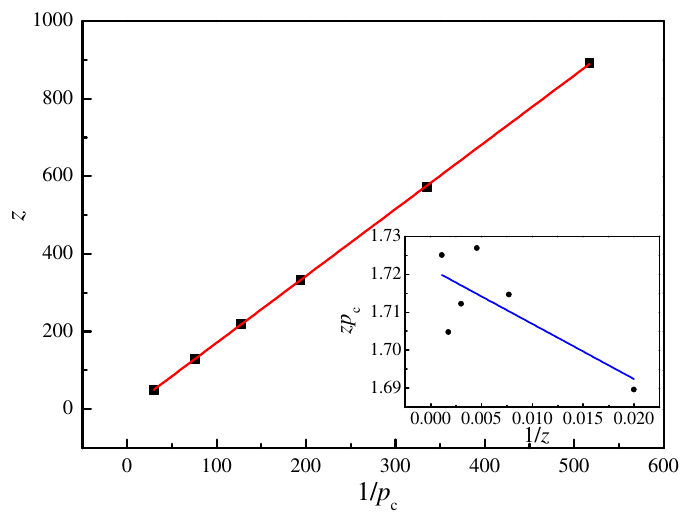}
\caption{
A plot of $z$ versus\ $1/p_c$ for site percolation on the \textsc{sc(5)}-$1...n$ lattices, and a fit $1.722/(z+1)$, implying $c = 1.722(7)$ and $b=1$ of Eq.\ (\ref{eq:zpcsite2}). Note that $c$ is somewhat less than the predicted value $32 \eta_c = 1.742$, reflecting either the relatively small systems considered, or possible a somewhat high value of $\eta_c$.  Inset: $z p_c$ versus $1/z$, showing rather larger scatter, presumably due to the differing shapes of the neighborhoods, and is not statistical error (error bars are smaller than the size of the symbols).  Line is the same fit as in (a).}
\label{fig:sitezpcb}
\end{figure}

For site percolation on lattices with compact nearest neighborhoods, the asymptotic behavior between thresholds $p_c$ and coordination number $z$ is expected to follow $zp_c \sim 32\eta_c = 1.742(2)$, where $\eta_c = 0.05443(7)$\cite{TorquatoJiao2012} is the continuum percolation threshold of five-dimensional hyperspheres. However, a plot of $z p_c$ versus $1/z$ shown in Fig.\ \ref{fig:sitezpcb} shows that our data is somewhat below this predicted value, and large fluctuations occur.  These fluctuations are not statistical in nature but presumably due to the somewhat different shape of the 5-dimensional neighborhoods as $n$ is increased. We fit the data roughly to $1.722/(z+1)$, implying $c = 1.722$ and $b=1$ of Eq.\ (\ref{eq:zpcsite2}), as shown Fig.\ \ref{fig:sitezpcb}.

\section{Conclusions}
\label{sec:conclusions}
To summarize, in this paper, by means of Monte Carlo simulations, both bond and site percolation on \textsc{sc(5)} lattice with extended neighborhoods are investigated based upon an effective single-cluster growth method. The main results are as follows.

(1) Simulations on the \textsc{sc(5)} lattices are carried out by using existing bond and site percolation thresholds, and the critical exponents of $\tau = 2.4177(3)$ and $\Omega = 0.27(2)$ are estimated by a self-consistent method. The result of $\tau$ is consistent with recent five-loop renormalization prediction of $2.4175(2)$ by Borinsky et al.\ \cite{BorinskyGraceyKompanietsSchnetz2021} as well as simulation results 2.419(1)  of Mertens and Moore \cite{MertensMoore2018} and 2.412(4)  of Paul, Ziff and Stanley \cite{PaulZiffStanley2001}.  It is also comparable to the value 2.4180(6) that follows from the result $d_f = 3.5260(14)$ of Zhang et al.\ \cite{ZhangHouFangHuDeng21}  and the hyperscaling relation $\tau = 1 + d/d_f$ with $d=5$.  The value of $\Omega$ is somewhat higher than the value 0.210(2) found in Ref.\ \cite{BorinskyGraceyKompanietsSchnetz2021} but in agreement with the central value with an older series analysis of Adler et al.\ \cite{AdlerMeirAharonyHarris90} which gave 0.27(7).

(2) For the bond percolation threshold of the \textsc{sc(5)} lattice, we find $p_c =  0.11817150(5)$, which is consistent with and confirms the precise value $p_c =  0.11817145(3)$ of Mertens and Moore \cite{MertensMoore2018}.

(3) For bond percolation on \textsc{sc(5)} lattices with extended neighborhoods, the asymptotic value of $zp_c$ tends to Bethe-lattice behavior, and the finite-$z$ correction is found to be consistent with  $zp_{c} - 1 \sim a_0 z^{-0.88}$.  This data could also be fitted with $zp_{c} - 1 \sim a_1(3+\ln z)/z$, similar to the form conjectured by Frei and Perkins in 4 dimensions.

(4) For site percolation on \textsc{sc(5)} lattice with extended neighborhoods, the asymptotic behavior can roughly be accounted for by $p_c \sim c/(z + b)$ with $c = 1.722$ and $b = 1$.  This value of $c$ is somewhat below the predicted value $c = 2^5 \eta_c = 1.742(2)$ using the result $\eta_c = 0.05433(7)$ \cite{TorquatoJiao2012}.  Using larger neighborhoods should give better agreement (the maximum radius here is just $\sqrt{7}$), or another possibility is that the value of $\eta_c$ given in Ref.\ \cite{TorquatoJiao2012} is somewhat high.  Resolving this would be an interesting area for future study.

\begin{figure}[htbp] 
\centering
\includegraphics[width=2.9in]{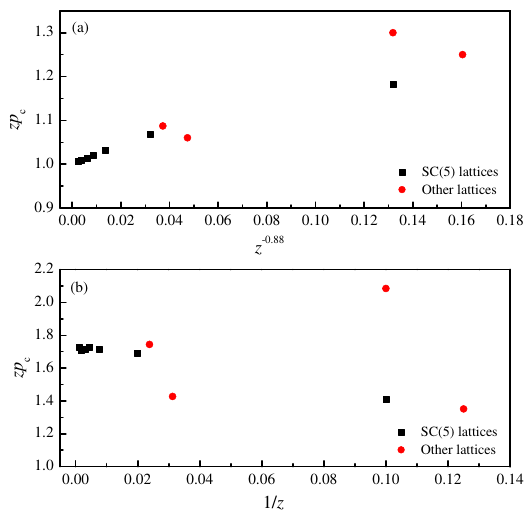}
\caption{Plots of $z p_c$ versus $z^{-0.88}$ for bond percolation (a) and versus $1/z$ for site percolation (b), showing results for the four systems listed in table  \ref{tab:otherlattices} (circles) along with the results for the  \textsc{sc(5)}-$1,...,n$ lattices (squares) as plotted in Figs.\ \ref{fig:bondzpc2} and \ref{fig:sitezpcb}.}
\label{fig:zpcbondsite}\
\end{figure}

(5)  For comparison with other (short-range) lattices in five dimensions, values of their $z p_c$ are given in table \ref{tab:otherlattices}.  {Plotting these as $z$ versus\ $1/p_c$ yields an apparently straight line continuing the behavior seen for example for site percolation in Fig.\ \ref{fig:sitezpcb}.  However, plotting $zp_c$ versus\ $z^{-0.88}$ for bond percolation and versus\ $1/z$ for site percolation, which amplifies the differences from $1/z$ behavior, shows more erratic behavior for these lattices (Fig.\ \ref{fig:zpcbondsite}).  These lattices do not in general fit the smooth $z$-dependent behavior that we have found for the lattices with the compact neighborhoods.  Note that two lattices that have the same value of $z=10$, the \textsc{sc(5)} and kagome(5) lattices, have fairly different thresholds.}

Some of these additional lattices can be mapped to the \textsc{sc(5)} lattice with extended bonds.  For example, the fcc lattice is equivalent to the \textsc{sc(5)} lattice with connections to just the second NN (\textsc{sc(5)}-0,1), and the bcc lattice is equivalent to \textsc{sc(5)} with connections to the 32 corners.

\begin{table}[htb]
\caption{Values of $z p_c$ of other five-dimensional lattices.}
\begin{tabular}{crllc}
\hline\hline
    lattice                   & $z$    & $z p_{c}$(site)    & $zp_{c}$(bond) & Refs.\\ \hline
    diamond          & 8     & 1.3512(18)     & 1.250(2)  &  \cite{vanderMarck1998,vanderMarck98a}  \\
    kagome           & 10    & 2.084(4)   & 1.30(2)   & \cite{vanderMarck1998,vanderMarck98a}  \\
    bcc      & 32    & 1.427(13)   & 1.06(3)   & \cite{vanderMarck1998,vanderMarck98a} \\ 
    fcc      & 40    & 1.74365(2)   & 1.087252(8)   &  \cite{HuCharbonneau2021}\\
\hline\hline
\end{tabular}
\label{tab:otherlattices}
\end{table}

Overall, percolation with extended-range bonds is interesting for both bond and site percolation, and has many connections with the literature in the field.

\section{Acknowledgments}
The authors thank Hao Hu, Youjin Deng, Stephan Mertens and John Gracey for helpful discussions.  This work is supported by National Natural Science Foundation of China (No.\ 12275002) and the Key Academic Discipline Project of China University of Mining and Technology (No.\ 2022WLXK01).

\bibliographystyle{unsrt}
\bibliography{bibliography.bib}

\begin{thebibliography}{10}

\bibitem{BroadbentHammersley1957}
S.~R. Broadbent and J.~M. Hammersley.
\newblock Percolation processes: {I}. {C}rystals and mazes.
\newblock {\em Math. Proc. Cambridge Phil. Soc.}, 53(3):629–641, 1957.

\bibitem{StaufferAharony1994}
Dietrich Stauffer and Amnon Aharony.
\newblock {\em {Introduction to Percolation Theory, 2nd revised ed.}}
\newblock Taylor and Francis, 1994.

\bibitem{BolandtabaSkauge2011}
S.~F. Bolandtaba and A.~Skauge.
\newblock Network modeling of {EOR} processes: A combined invasion percolation
  and dynamic model for mobilization of trapped oil.
\newblock {\em Transport in Porous Media}, 89(3):357--382, 2011.

\bibitem{MourzenkoThovertAdler2011}
V.~V. Mourzenko, J.-F. Thovert, and P.~M. Adler.
\newblock Permeability of isotropic and anisotropic fracture networks, from the
  percolation threshold to very large densities.
\newblock {\em Phys. Rev. E}, 84:036307, 2011.

\bibitem{Henley1993}
Christopher~L. Henley.
\newblock Statics of a ``self-organized'' percolation model.
\newblock {\em Phys. Rev. Lett.}, 71:2741--2744, 1993.

\bibitem{GuisoniLoscarAlbano2011}
Nara Guisoni, Ernesto~S. Loscar, and Ezequiel~V. Albano.
\newblock Phase diagram and critical behavior of a forest-fire model in a
  gradient of immunity.
\newblock {\em Phys. Rev. E}, 83:011125, 2011.

\bibitem{RahmanHassan19}
M.~S. Rahman and M.~K. Hassan.
\newblock Redefinition of site percolation in light of entropy and the second
  law of thermodynamics.
\newblock {\em Phys. Rev. E}, 100:062109, 2019.

\bibitem{MooreNewman2000}
Cristopher Moore and M.~E.~J. Newman.
\newblock Epidemics and percolation in small-world networks.
\newblock {\em Phys. Rev. E}, 61:5678--5682, 2000.

\bibitem{ScullardJacobsen2020}
Christian~R. Scullard and Jesper~Lykke Jacobsen.
\newblock Bond percolation thresholds on {A}rchimedean lattices from critical
  polynomial roots.
\newblock {\em Phys. Rev. Res.}, 2:012050, 2020.

\bibitem{Ziff2021}
Robert~M. Ziff.
\newblock Percolation and the pandemic.
\newblock {\em Physica A}, 568:125723, 2021.

\bibitem{XuWangHuDeng2021}
Wenhui Xu, Junfeng Wang, Hao Hu, and Youjin Deng.
\newblock Critical polynomials in the nonplanar and continuum percolation
  models.
\newblock {\em Phys. Rev. E}, 103:022127, 2021.

\bibitem{MertensMoore2012}
Stephan Mertens and Cristopher Moore.
\newblock Continuum percolation thresholds in two dimensions.
\newblock {\em Phys. Rev. E}, 86:061109, 2012.

\bibitem{QuintanillaZiff2007}
John~A. Quintanilla and Robert~M. Ziff.
\newblock Asymmetry in the percolation thresholds of fully penetrable disks
  with two different radii.
\newblock {\em Phys. Rev. E}, 76:051115, 2007.

\bibitem{TarasevichEserkepov20}
Yuri~Yu.\ Tarasevich and Andrei~V. Eserkepov.
\newblock Percolation thresholds for discorectangles: Numerical estimation for
  a range of aspect ratios.
\newblock {\em Phys. Rev. E}, 101:022108, 2020.

\bibitem{Kantor1986}
Yacov Kantor.
\newblock Three-dimensional percolation with removed lines of sites.
\newblock {\em Phys. Rev. B}, 33:3522--3525, 1986.

\bibitem{ZierenbergFrickeMarenzSpitznerBlavatskaJanke2017}
Johannes Zierenberg, Niklas Fricke, Martin Marenz, F.~P. Spitzner, Viktoria
  Blavatska, and Wolfhard Janke.
\newblock Percolation thresholds and fractal dimensions for square and cubic
  lattices with long-range correlated defects.
\newblock {\em Phys. Rev. E}, 96:062125, 2017.

\bibitem{ChoiYu2020}
Jeong-Ok Choi and Unjong Yu.
\newblock Bootstrap and diffusion percolation transitions in three-dimensional
  lattices.
\newblock {\em J. Stat. Mech: Theor. Exp.}, 2020(6):063218, 2020.

\bibitem{ChoiYu2019}
Jeong-Ok Choi and Unjong Yu.
\newblock Newman-{Z}iff algorithm for the bootstrap percolation: Application to
  the {A}rchimedean lattices.
\newblock {\em J. Comput. Phys.}, 386:1--8, 2019.

\bibitem{MuroBuldyrevBraunstein2020}
Mat\'{\i}as~A. Di~Muro, Sergey~V. Buldyrev, and Lidia~A. Braunstein.
\newblock Reversible bootstrap percolation: Fake news and fact checking.
\newblock {\em Phys. Rev. E}, 101:042307, 2020.

\bibitem{MuroValdezStanleyBuldyrevBraunstein2019}
Mat\'{\i}as~A. Di~Muro, Lucas~D. Valdez, H.~Eugene Stanley, Sergey~V. Buldyrev,
  and Lidia~A. Braunstein.
\newblock Insights into bootstrap percolation: Its equivalence with k-core
  percolation and the giant component.
\newblock {\em Phys. Rev. E}, 99:022311, 2019.

\bibitem{WangZhouLiuGaroniDeng13}
Junfeng Wang, Zongzheng Zhou, Qingquan Liu, Timothy~M. Garoni, and Youjin Deng.
\newblock High-precision {Monte} {C}arlo study of directed percolation in (d +
  1) dimensions.
\newblock {\em Phys. Rev. E}, 88:042102, 2013.

\bibitem{Grassberger2009}
Peter Grassberger.
\newblock Logarithmic corrections in $(4+1)$-dimensional directed percolation.
\newblock {\em Phys. Rev. E}, 79:052104, 2009.

\bibitem{Grassberger2009b}
Peter Grassberger.
\newblock Local persistence in directed percolation.
\newblock {\em J. Stat. Mech.: Th. Exp.}, 2009(08):P08021, 2009.

\bibitem{Jensen2004}
Iwan Jensen.
\newblock Low-density series expansions for directed percolation: {III}. {S}ome
  two-dimensional lattices.
\newblock {\em J. Phys. A: Math. Gen.}, 37(27):6899--6915, 2004.

\bibitem{KozaKondratSuszcaynski2014}
Zbigniew Koza, Grzegorz Kondrat, and Karol Suszczy{\'{n}}ski.
\newblock Percolation of overlapping squares or cubes on a lattice.
\newblock {\em J. Stat. Mech.: Th. Exp.}, 2014:P11005, 2014.

\bibitem{KozaPola2016}
Zbigniew {Koza} and Jakub {Po{\l}a}.
\newblock {From discrete to continuous percolation in dimensions 3 to 7}.
\newblock {\em J. Stat. Mech.: Th. Exp.}, 10:103206, 2016.

\bibitem{Kleinberg2000}
Jon~M. Kleinberg.
\newblock Navigation in a small world.
\newblock {\em Nature}, 406:845, 2000.

\bibitem{SanderWarrenSokolov2003}
L.~M. Sander, C.~P. Warren, and I.~M. Sokolov.
\newblock Epidemics, disorder, and percolation.
\newblock {\em Physica A}, 325(1):1 -- 8, 2003.

\bibitem{DaltonDombSykes64}
N.~W. Dalton, C.~Domb, and M.~F. Sykes.
\newblock Dependence of critical concentration of a dilute ferromagnet on the
  range of interaction.
\newblock {\em Proc. Phys. Soc.}, 83(3):496--498, 1964.

\bibitem{DombDalton1966}
C.~Domb and N.~W. Dalton.
\newblock Crystal statistics with long-range forces: I. {T}he equivalent
  neighbour model.
\newblock {\em Proc. Phys. Soc.}, 89(4):859--871, 1966.

\bibitem{Domb72}
C.~Domb.
\newblock A note on the series expansion method for clustering problems.
\newblock {\em Biometrika}, 59(1):209–211, 1972.

\bibitem{GoukerFamily83}
Mark Gouker and Fereydoon Family.
\newblock Evidence for classical critical behavior in long-range site
  percolation.
\newblock {\em Phys. Rev. B}, 28:1449--1452, 1983.

\bibitem{JerauldScrivenDavis1984}
G.~R. Jerauld, L.~E. Scriven, and H.~T. Davis.
\newblock Percolation and conduction on the 3d {V}oronoi and regular networks:
  a second case study in topological disorder.
\newblock {\em J. Phys. C: Solid State}, 17(19):3429--3439, 1984.

\bibitem{GawronCieplak91}
T.~R. Gawron and Marek Cieplak.
\newblock Site percolation thresholds of fcc lattice.
\newblock {\em Acta Phys. Pol. A}, 80(3):461--464, 1991.

\bibitem{dIribarneRasigniRasigni95}
C.~d'Iribarne, G.~Rasigni, and M.~Rasigni.
\newblock Determination of site percolation transitions for 2d mosaics by means
  of the minimal spanning tree approach.
\newblock {\em Phys. Lett. A}, 209(1):95--98, 1995.

\bibitem{dIribarneRasigniRasigni99}
C.~d'Iribarne, M.~Rasigni, and G.~Rasigni.
\newblock Minimal spanning tree and percolation on mosaics: {G}raph theory and
  percolation.
\newblock {\em J. Phys. A: Math. Gen.}, 32(14):2611--2622, 1999.

\bibitem{dIribarneRasigniRasigni99b}
C.~d'Iribarne, M.~Rasigni, and G.~Rasigni.
\newblock From lattice long-range percolation to the continuum one.
\newblock {\em Phys. Lett. A}, 263(1):65--69, 1999.

\bibitem{MalarzGalam05}
Krzysztof Malarz and Serge Galam.
\newblock Square-lattice site percolation at increasing ranges of neighbor
  bonds.
\newblock {\em Phys. Rev. E}, 71:016125, 2005.

\bibitem{GalamMalarz05}
Serge Galam and Krzysztof Malarz.
\newblock Restoring site percolation on damaged square lattices.
\newblock {\em Phys. Rev. E}, 72:027103, 2005.

\bibitem{MajewskiMalarz2007}
M.~Majewski and K.~Malarz.
\newblock Square lattice site percolation thresholds for complex
  neighbourhoods.
\newblock {\em Acta Phys. Pol. B}, 38:2191, 2007.

\bibitem{KurzawskiMalarz2012}
{\L}ukasz Kurzawski and Krzysztof Malarz.
\newblock Simple cubic random-site percolation thresholds for complex
  neighbourhoods.
\newblock {\em Rep. Math. Phys.}, 70(2):163--169, 2012.

\bibitem{Malarz2015}
Krzysztof Malarz.
\newblock Simple cubic random-site percolation thresholds for neighborhoods
  containing fourth-nearest neighbors.
\newblock {\em Phys. Rev. E}, 91:043301, 2015.

\bibitem{KotwicaGronekMalarz19}
M.~Kotwica, P.~Gronek, and K.~Malarz.
\newblock Efficient space virtualization for the {H}oshen-{K}opelman algorithm.
\newblock {\em Int. J. Mod. Phys. C}, 30(8):1950055, 2019.

\bibitem{Malarz2020}
Krzysztof Malarz.
\newblock Site percolation thresholds on triangular lattice with complex
  neighborhoods.
\newblock {\em Chaos}, 30:123123, 2020.

\bibitem{Malarz21}
Krzysztof Malarz.
\newblock Percolation thresholds on a triangular lattice for neighborhoods
  containing sites up to the fifth coordination zone.
\newblock {\em Phys. Rev. E}, 103:052107, 2021.

\bibitem{Malarz23}
Krzysztof Malarz.
\newblock Random site percolation thresholds on square lattice for complex
  neighborhoods containing sites up to the sixth coordination zone.
\newblock {\em preprint arXiv 2303.10423}, 2023.

\bibitem{MitraSahaSensharma22}
Sayantan Mitra, Dipa Saha, and Ankur Sensharma.
\newblock Percolation in a simple cubic lattice with distortion.
\newblock {\em Phys. Rev. E}, 106:034109, 2022.

\bibitem{MitraSensharma23}
Sayantan Mitra and Ankur Sensharma.
\newblock Site percolation in distorted square and simple cubic lattices with
  flexible number of neighbors.
\newblock {\em Phys. Rev. E}, 107:064127, 2023.

\bibitem{OuyangDengBlote2018}
Yunqing Ouyang, Youjin Deng, and Henk W.~J. Bl\"ote.
\newblock Equivalent-neighbor percolation models in two dimensions: Crossover
  between mean-field and short-range behavior.
\newblock {\em Phys. Rev. E}, 98:062101, 2018.

\bibitem{DengOuyangBlote2019}
Youjin Deng, Yunqing Ouyang, and Henk W.~J. Blöte.
\newblock Medium-range percolation in two dimensions.
\newblock {\em J. Phys.: Conf. Ser.}, 1163:012001, 2019.

\bibitem{XunZiff2020}
Zhipeng Xun and Robert~M. Ziff.
\newblock Precise bond percolation thresholds on several four-dimensional
  lattices.
\newblock {\em Phys. Rev. Research}, 2:013067, 2020.

\bibitem{XunZiff2020b}
Zhipeng Xun and Robert~M. Ziff.
\newblock Bond percolation on simple cubic lattices with extended
  neighborhoods.
\newblock {\em Phys. Rev. E}, 102:012102, 2020.

\bibitem{XunHaoZiff2022}
Zhipeng Xun, Dapeng Hao, and Robert~M. Ziff.
\newblock Site and bond percolation thresholds on regular lattices with compact
  extended-range neighborhoods in two and three dimensions.
\newblock {\em Phys. Rev. E}, 105:024105, 2022.

\bibitem{ZhaoYanXunHaoZiff2022}
Pengyu Zhao, Jinhong Yan, Zhipeng Xun, Dapeng Hao, and Robert~M Ziff.
\newblock Site and bond percolation on four-dimensional simple hypercubic
  lattices with extended neighborhoods.
\newblock {\em J. Stat. Mech.: Th. Exp.}, 2022(3):033202, 2022.

\bibitem{FreiPerkins2016}
Spencer Frei and Edwin Perkins.
\newblock A lower bound for $p_c$ in range-$r$ bond percolation in two and
  three dimensions.
\newblock {\em Electron. J. Probab.}, 21:56, 2016.

\bibitem{Hong21}
Jieliang Hong.
\newblock An upper bound for $p_c$ in range-$r$ bond percolation in two and
  three dimensions.
\newblock {\em preprint arXiv 2107.14173}, 2021.

\bibitem{Penrose93}
Mathew~D. Penrose.
\newblock {On the spread-out limit for bond and continuum percolation}.
\newblock {\em The Annals of Applied Probability}, 3(1):253--276, 1993.

\bibitem{XunHaoZiff2021}
Zhipeng Xun, Dapeng Hao, and Robert~M. Ziff.
\newblock Site percolation on square and simple cubic lattices with extended
  neighborhoods and their continuum limit.
\newblock {\em Phys. Rev. E}, 103:022126, 2021.

\bibitem{MertensMoore2018}
Stephan Mertens and Cristopher Moore.
\newblock Percolation thresholds and {F}isher exponents in hypercubic lattices.
\newblock {\em Phys. Rev. E}, 98:022120, 2018.

\bibitem{JanHongStanley85}
N.~Jan, D.~C. Hong, and H.~E. Stanley.
\newblock The fractal dimension and other percolation exponents in four and
  five dimensions.
\newblock {\em J. Phys. A}, 18(15):L935, 1985.

\bibitem{KnackstedtMcCraryPayandehRoberts88}
M~Knackstedt, J~McCrary, B~Payandeh, and M~Robert.
\newblock Block cluster theory of site percolation on the four- and
  five-dimensional ordinary hypercubic lattices.
\newblock {\em Journal of Physics A: Mathematical and General}, 21(21):4067,
  nov 1988.

\bibitem{LorenzZiff1998}
Christian~D. Lorenz and Robert~M. Ziff.
\newblock Precise determination of the bond percolation thresholds and
  finite-size scaling corrections for the sc, fcc, and bcc lattices.
\newblock {\em Phys. Rev. E}, 57:230--236, 1998.

\bibitem{ZiffBabalievski99}
Robert~M. Ziff and Filip Babalievski.
\newblock Site percolation on the {P}enrose rhomb lattice.
\newblock {\em Physica A}, 269:201--210, 1999.

\bibitem{Ziff98}
Robert~M. Ziff.
\newblock Four-tap shift-register-sequence random-number generators.
\newblock {\em Computer in Physics}, 12(4):385--392, 1998.

\bibitem{Kirkpatrick1976}
Scott Kirkpatrick.
\newblock Percolation phenomena in higher dimensions: Approach to the
  mean-field limit.
\newblock {\em Phys. Rev. Lett.}, 36:69--72, 1976.

\bibitem{GauntSykesRuskin1976}
D.~S. Gaunt, M.~F. Sykes, and H.~Ruskin.
\newblock Percolation processes in d-dimensions.
\newblock {\em J. Phys. A: Math. Gen.}, 9(11):1899--1911, 1976.

\bibitem{Grassberger2003}
Peter Grassberger.
\newblock Critical percolation in high dimensions.
\newblock {\em Phys. Rev. E}, 67:036101, 2003.

\bibitem{DammerHinrichsen2004}
Stephan~M. {Dammer} and Haye {Hinrichsen}.
\newblock {Spreading with immunization in high dimensions}.
\newblock {\em J. Stat. Mech.: Th. Exp.}, 7:07011, 2004.

\bibitem{MertensMoore2018b}
Stephan Mertens and Cristopher Moore.
\newblock Series expansion of the percolation threshold on hypercubic lattices.
\newblock {\em J. Phys. A}, 51(47):475001, oct 2018.

\bibitem{AdlerMeirAharonyHarris90}
Joan Adler, Yigal Meir, Amnon Aharony, and A.~B. Harris.
\newblock Series study of percolation moments in general dimension.
\newblock {\em Phys. Rev. B}, 41:9183--9206, 1990.

\bibitem{BorinskyGraceyKompanietsSchnetz2021}
M.~Borinsky, J.~A. Gracey, M.~V. Kompaniets, and O.~Schnetz.
\newblock Five-loop renormalization of ${\ensuremath{\phi}}^{3}$ theory with
  applications to the {L}ee-{Y}ang edge singularity and percolation theory.
\newblock {\em Phys. Rev. D}, 103:116024, 2021.

\bibitem{ZhangHouFangHuDeng21}
Zhongjin Zhang, Pengcheng Hou, Sheng Fang, Hao Hu, and Youjin Deng.
\newblock Critical exponents and universal excess cluster number of percolation
  in four and five dimensions.
\newblock {\em Physica A}, 580:126124, 2021.

\bibitem{TanDengJacobsen20}
Xiao-Jun Tan, You-Jin Deng, and Jesper~Lykke Jacobsen.
\newblock N-cluster correlations in four- and five-dimensional percolation.
\newblock {\em Frontiers of Physics}, 15:41501, 2020.

\bibitem{PaulZiffStanley2001}
Gerald Paul, Robert~M. Ziff, and H.~Eugene Stanley.
\newblock Percolation threshold, {F}isher exponent, and shortest path exponent
  for four and five dimensions.
\newblock {\em Phys. Rev. E}, 64:026115, 2001.

\bibitem{Gracey2015}
J.~A. Gracey.
\newblock Four loop renormalization of ${\ensuremath{\phi}}^{3}$ theory in six
  dimensions.
\newblock {\em Phys. Rev. D}, 92:025012, 2015.

\bibitem{TorquatoJiao2012}
S.~Torquato and Y.~Jiao.
\newblock Effect of dimensionality on the continuum percolation of overlapping
  hyperspheres and hypercubes. {II.} {S}imulation results and analyses.
\newblock {\em J. Chem. Phys.}, 137:074106, 2012.

\bibitem{vanderMarck1998}
Steven~C. {van der Marck}.
\newblock Calculation of percolation thresholds in high dimensions for fcc, bcc
  and diamond lattices.
\newblock {\em Int. J. Mod. Phys. C}, 9(4):529--540, 1998.

\bibitem{vanderMarck98a}
Steven~C. van~der Marck.
\newblock Site percolation and random walks on d-dimensional kagom{\'{e}}
  lattices.
\newblock {\em J. Phys. A: Math. Gen.}, 31(15):3449--3460, 1998.

\bibitem{HuCharbonneau2021}
Yi~Hu and Patrick Charbonneau.
\newblock Percolation thresholds on high-dimensional ${D}_{n}$ and
  ${E}_{8}$-related lattices.
\newblock {\em Phys. Rev. E}, 103:062115, 2021.

\end{thebibliography}

\end{document}